\documentclass[10pt]{article}
\usepackage{moriond,epsfig}

\begin{document}
\vspace*{3cm}
\title{LATEST RESULTS OF THE EDELWEISS EXPERIMENT}

\author{ V. SANGLARD for the Edelweiss collaboration~\cite{edel}}

\address{Institut de Physique Nucl\'eaire de Lyon, 4 rue Enrico Fermi,\\
69622 Villeurbanne, France}

\maketitle\abstracts{
The Edelweiss experiment is a direct detection Dark Matter Search, under the form 
of WIMPs. It uses heat and ionization Ge cryogenic detectors.We present 
the latest results obtained by the experiment with three new 320~g 
bolometers. At present, Edelweiss I is the most sensitive experiment for all
WIMP masses compatible with accelerator constraints ($M_{WIMP}~>$~30~GeV).We 
also briefly describe the status of the second stage Edelweiss-II involving
initially 10~kg of detectors and aiming a gain of two orders of
magnitude in sensitivity.}

\section{Introduction}
Recent cosmological observations~\cite{{WMAP},{WMAP2}} of the CMB show that the main part of the 
matter in our Universe is dark and non baryonic. If non baryonic Dark Matter is made of particles,
they must be stable, neutral and massive : WIMPs (Weakly Interactive Masssive Particles). 
In the MSSM (Minimal Supersymmetric Standard Model) framework, the WIMP could be the LSP 
(Lightest Supersymmetric Particle) called neutralino. It has a mass between few tens and 
few hundreds of GeV/c$^2$, and a scattering cross section with a nucleon below $10^{-6}$~pb.\\  
There are two methods to detect non baryonic Dark Matter under the form of WIMPs : 
direct, looking for interactions in a detector, and indirect, 
looking for annihilation products of WIMPs (for a complete review of WIMPs and
neutralino Dark Matter, see~\cite{jungman}).\\
The Edelweiss experiment is dedicated to the direct detection like other experiments
(DAMA~\cite{dama},CDMS~\cite{cdms}, ZEPLIN~\cite{zeplin}, CRESST~\cite{cresst}). 

\section{The Edelweiss experiment}

\subsection{Direct detection}

In this technique, a WIMP is detected by measuring the nuclear recoil produced by its 
elastic interaction with an ordinary matter target. \\
The background consists of two different contributions : neutrons that produce nuclear recoils like 
WIMPs, and electrons or gammas that induce electron recoils in the target.\\
The WIMP properties from the most optimistic SUSY models lead to strong constraints 
on experiments. Indeed calculations predict a deposited energy typically below 100~keV and
an interaction rate with ordinary matter below 1~evt/kg/day, requiring experiments with
 a very low background and large detector masses.
 
\subsection{The experimental setup}

The Edelweiss experiment, described elsewhere~\cite{{edel1},{edel2}}, is
located at the LSM (french acronym for Modane Underground Laboratory) adjacent to the
Frejus tunnel connecting France and Italy. About 1700 m of rock protect the experiment 
from radioactive background generated by cosmic rays. In the 
laboratory, the muon flux is reduced by a factor $2\times 10^6$ compared to the flux
at sea level. The neutron flux from the rock radioactivity has been 
measured~\cite{{verene},{neutron}} to be $\sim 1.6\times 10^{-6}$~cm$^2$.s$^{-1}$. The experiment 
is surrounded by passive shielding made of paraffin (30 cm), lead (15 cm) and copper (10 cm).\\
The Edelweiss experiment used cryogenic Ge bolometers described in more 
details in~\cite{{edel3},{edel4}}. They have a cylindrical geometry. To improve electric 
field geometry in the crystal, their edges are beveled at an angle of 45 degrees. 
One of the main characteristics of these detectors is the two simultaneous measurement of 
ionization and heat. The ionization signal is collected by aluminum electrodes on each 
side of the crystal. The tiny rise in temperature is measured by a NTD heat sensor glued 
onto one electrode. Moreover some detectors were equipped with an amorphous layer 
(in Ge or Si) to improve charge collection efficiency for near surface events~\cite{surface}.\\
On one side, the electrode is segmented in order to define a central fiducial volume and 
a guard ring. Most of the radioactivity due to the detector environment is collected on 
the latter part of the detector. The fiducial inner volume, defined as $\geq$~75$\%$ of 
the charge collected on the central electrode, corresponds to 57~$\%$ of the total detector 
volume~\cite{edel4}.\\
Since January 2002, three 320~g detectors have been simultaneously operated in a 
dilution cryostat working at a regulated temperature of 17~mK. To decrease the 
background, all materials around the detectors were carefully selected for their low
radioactivity. Therefore the front end electronic components are placed behind a roman 
lead shielding above the three detectors. With these precautions combined with the LSM experimental 
site quality and with the passive shielding surrounding the experiment~\cite{lsm}, 
the gamma ray background is only of $\sim$~1.5~evts/(keV.kg.day) between 20 and 100~keV, 
before the gamma rejection. Then residual neutron background is estimated to be 
0.03 evts/(kg.day) above 20~keV~\cite{neutron}. 
   
\subsection{Calibrations}

The heat and ionization responses to gamma rays were calibrated using $^{57}$Co and 
$^{137}$Cs radioactive sources. A summary of the bolometer baselines, resolutions 
and energy thresholds can be found in~\cite{edel4}.\\
The simultaneous measurement of both heat and ionization signals provides an excellent 
event by event discrimination between nuclear and electron recoils. The ratio
of the ionization and heat signals depends on the recoiling particle, since a nucleus
produces less ionization in crystal than an electron does.\\
During a $^{252}$Cf calibration run, the energy threshold for event by event 
discrimination can be obtained. Typically with the Edelweiss detectors it is possible
to reject more than 99.9~$\%$ of electron recoils down to 15~keV.
The rejection efficiency is a fundamental parameter for these types of detectors. 
It is regularly controlled by measuring the ratio of ionization to recoil energy during
gamma rays calibrations, where in addition, the charge collection quality for gamma 
rays can be checked.

\section{Latest results}
\subsection{Ionization trigger}
\label{data}

\begin{figure*}[!bht]
\begin{center}
  \resizebox{17cm}{!}{ 
\includegraphics[width=0.45\textwidth]{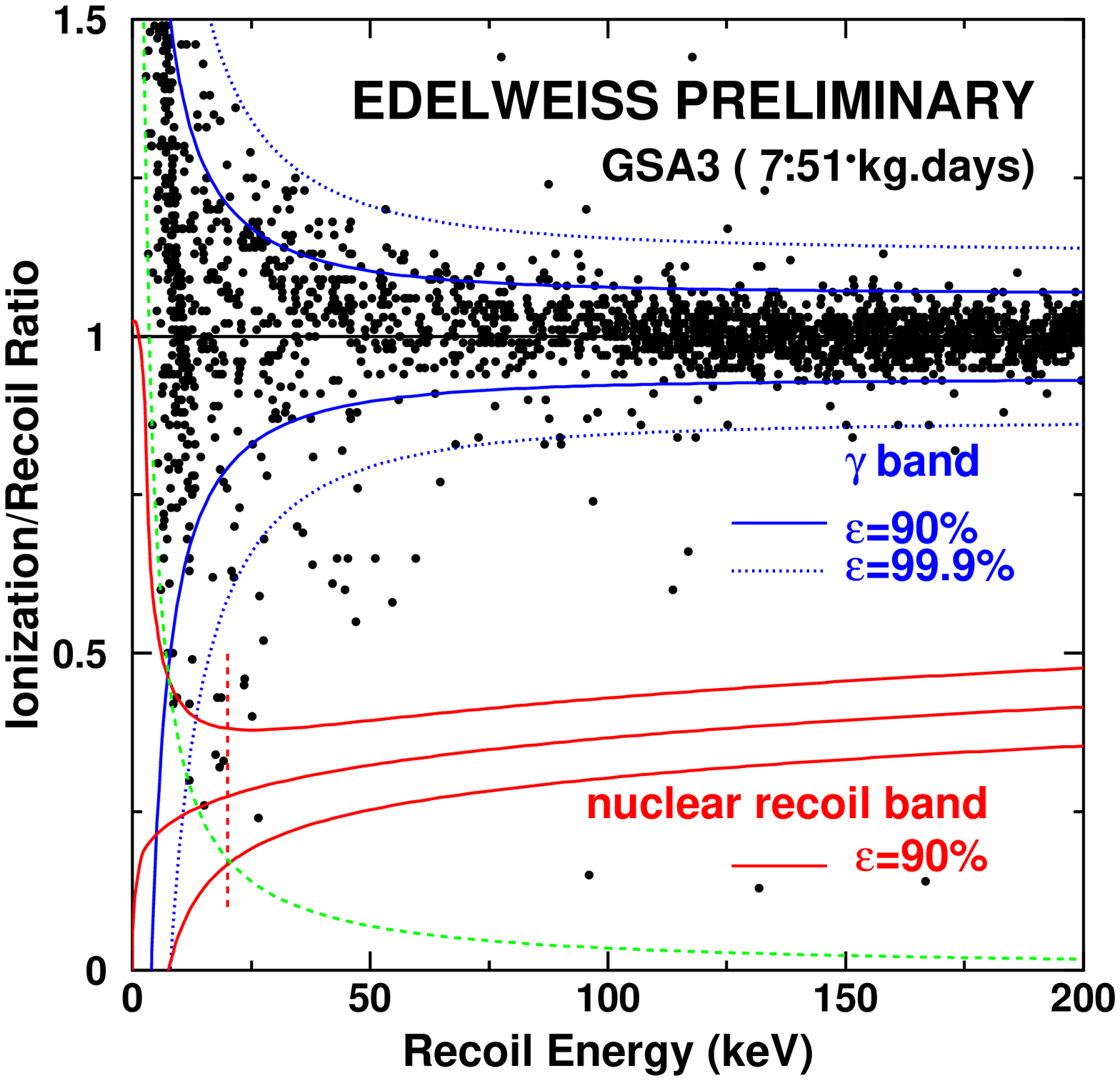}
\includegraphics[width=0.45\textwidth]{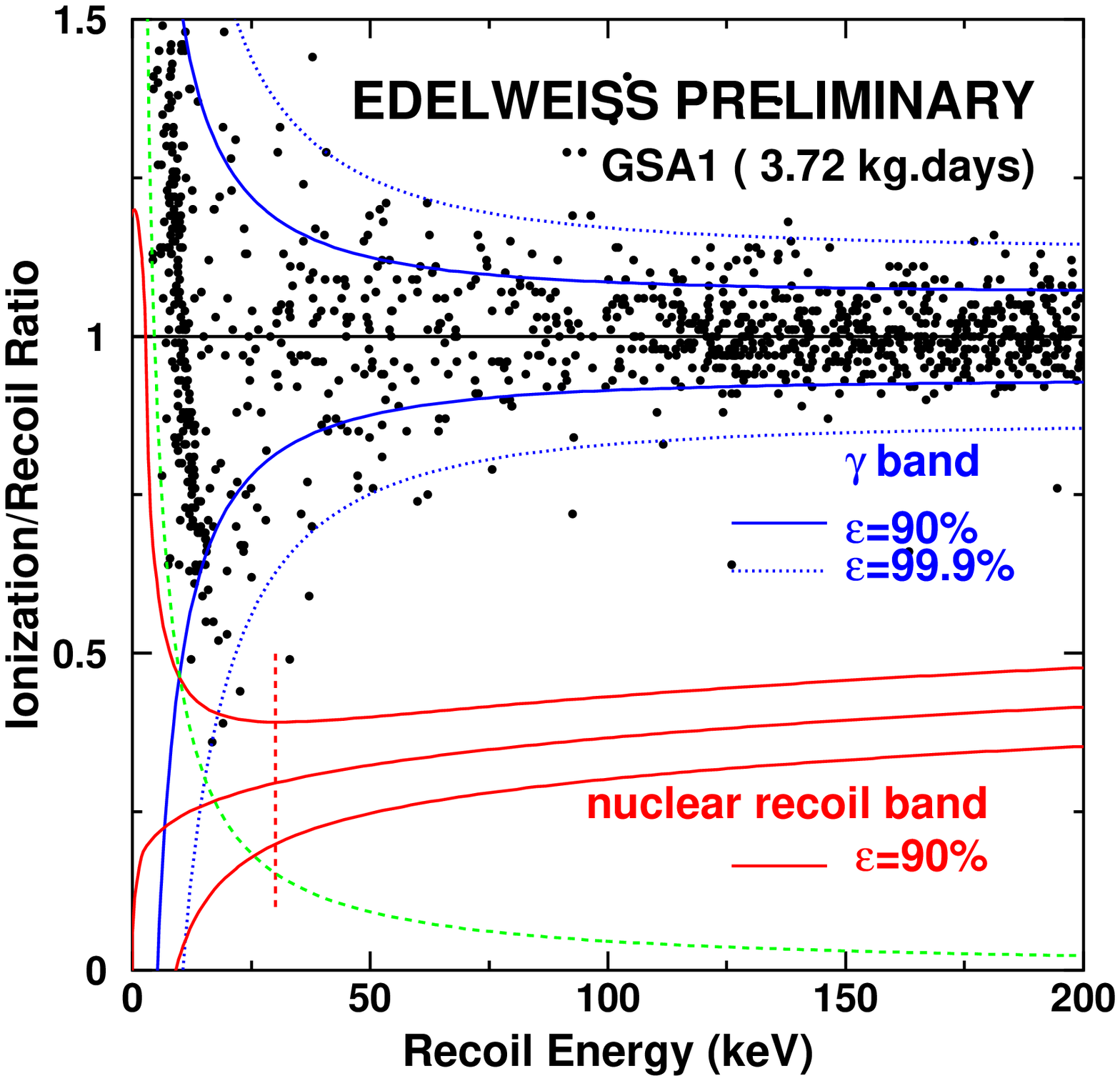}  
\includegraphics[width=0.45\textwidth]{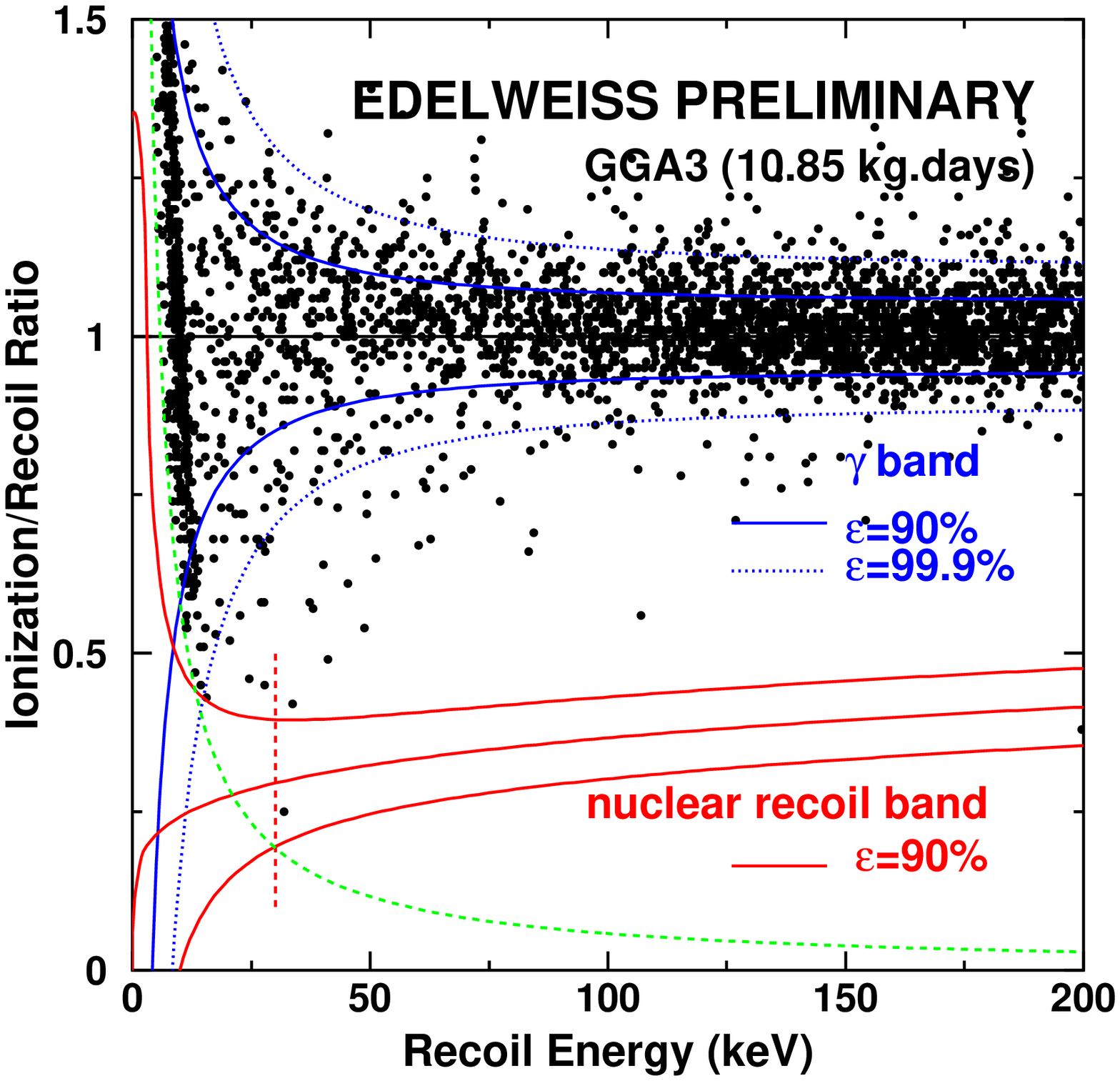} }
\caption{\footnotesize{$\frac{E_I}{E_R}$ versus $E_R$ (fiducial volume) for physics runs. The
nuclear recoil band is determined by the condition of 90~$\%$ acceptance.}}
\label{fond}
\end{center}
\end{figure*}

In 2000 and 2002, 11.6~kg.d were recorded with two different detectors~\cite{{edel},{edel1}}.
In 2003, three new detectors were placed in the cryostat and 20 kg.d were added to the previous 
published data. The recorded ratio of ionization over recoil energy as a function 
of recoil energy is represented in Fig.~\ref{fond}. The detectors have a similar behavior compatible to the previous 
data~\cite{{edel},{edel1}}.   

\begin{figure}[!bht]
\centering\epsfig{file=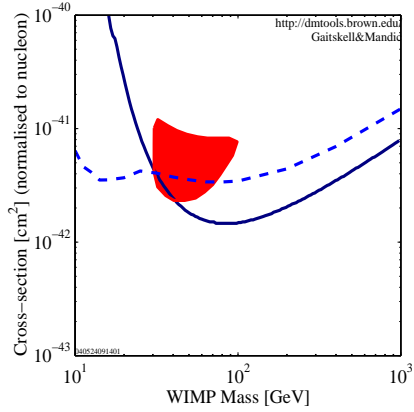,width=5.5cm,height=5.5cm}
\caption{\footnotesize{Preliminary exclusion limit (full line) for Edelweiss without 
background subtraction. Dashed curve : CDMS limit with background subtraction. Closed 
contour : allowed region at 3$\sigma$ C.L. from DAMA NaI1-4 annual modulation data.}}
\label{limit} 
\end{figure}

Three events compatible with nuclear recoils have been recorded. One of them has a recoil 
energy of about 200~keV, incompatible with a WIMP mass below 1~TeV/c$^2$. Due to the low 
statistics, the two other events are used to establish a 90~$\%$ C.L. upper 
limit on the WIMP-nucleon cross section as a function of the WIMP mass, shown in the fig.~\ref{limit}.

With no background subtraction, the new limit is identical to the previous one although it 
has been obtained with three new detectors and an extended exposure. The Edelweiss experiment 
has the best published limit for a WIMP mass above 30~GeV/c$^2$. This result confirms 
the incompatibility with a C.L. better than 99.8~$\%$ with the DAMA candidate~\cite{dama2} 
with a WIMP mass of about 44~GeV/c$^2$.

\subsection{Phonon trigger}

\begin{figure*}[!bht]
\begin{center}
 \resizebox{17cm}{!}{ 
\includegraphics[width=0.45\textwidth]{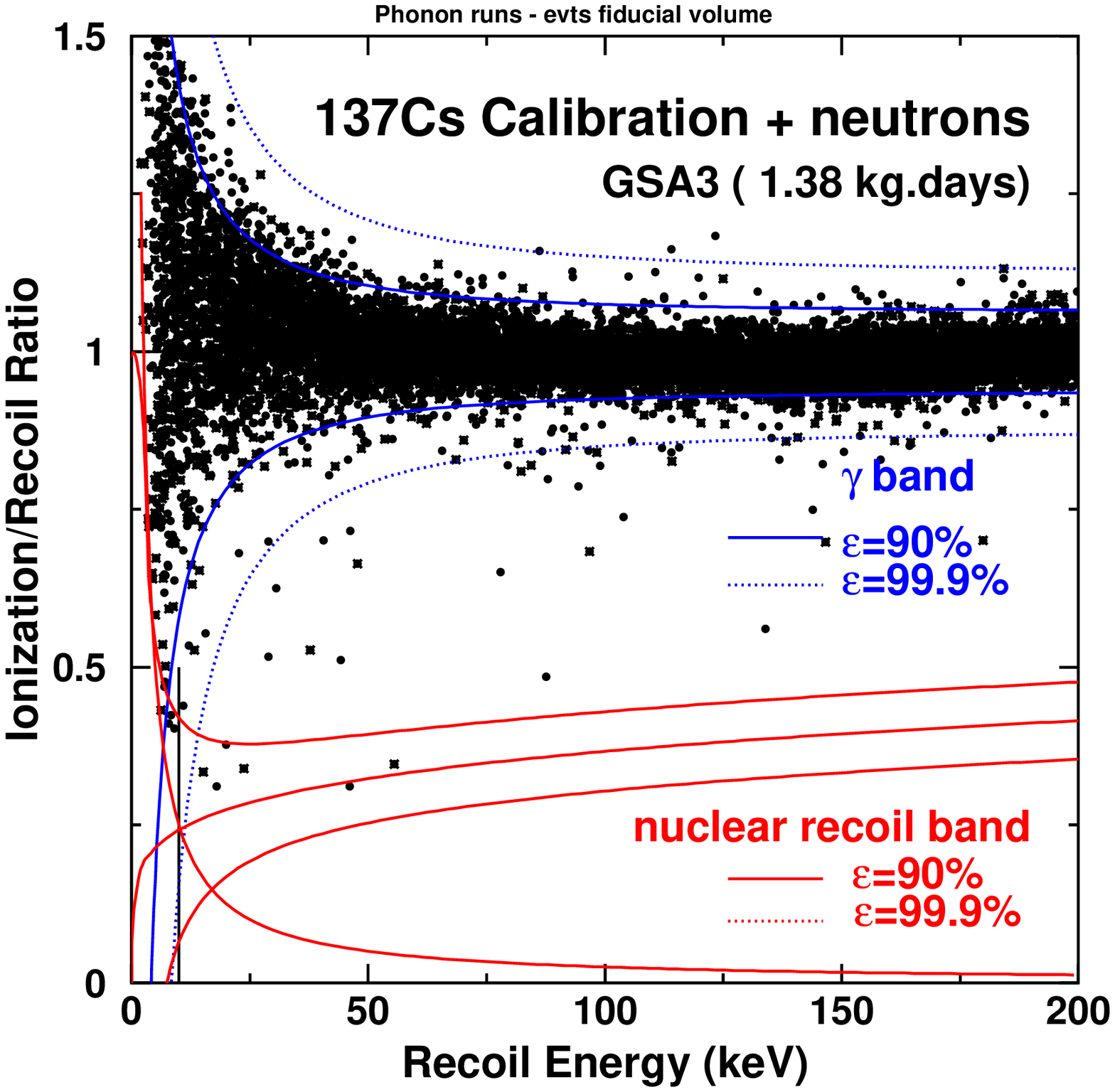}
\includegraphics[width=0.45\textwidth]{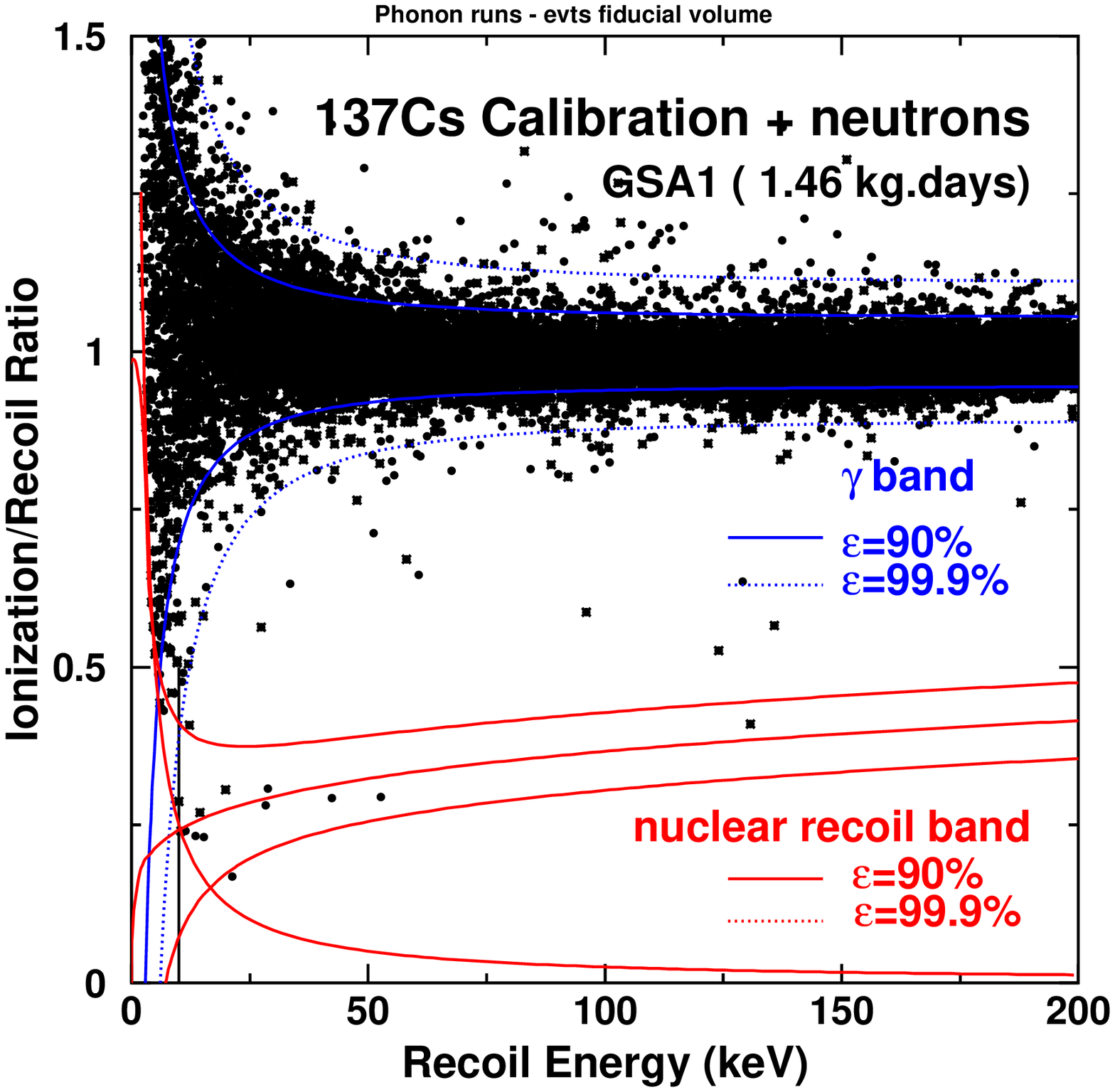}  
\includegraphics[width=0.45\textwidth]{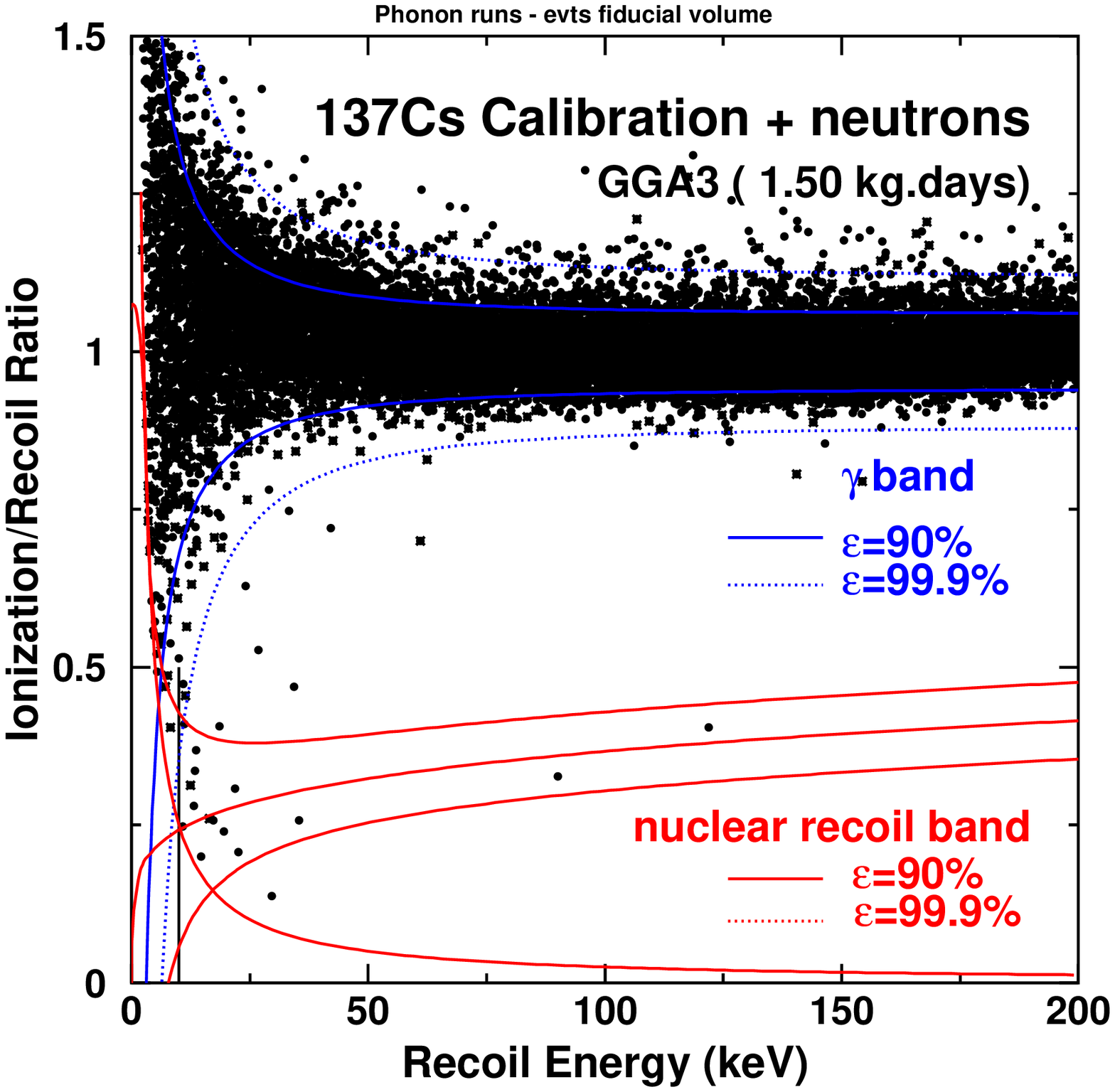} }
\caption{\footnotesize{$\frac{E_I}{E_R}$ versus $E_R$ (fiducial volume) for $^{137}$Cs calibration 
runs}}
\label{cesium}
\end{center}
\end{figure*}

After these results, the data taking continued with an improved trigger efficiency at low energy. 
Previously the trigger was the fast ionization signal, now the trigger is the phonon signal.
Thanks to a better resolution and no quenching factor on the phonon signal, a 100~$\%$ efficiency has 
been reached down to 10 and 15~keV depending on the detector.\\
During physics runs, calibrations must be made regularly. A high statistics gamma calibration is 
realized during the run, in order to check the charge collection quality and test the 
electron-nuclear recoil discrimination.\\
A two weeks gamma calibration with a  $^{137}$Cs source has been made equivalent to about two years
of physics run.
During this calibration, totalizing about 10$^5$ events, 31 
events (above 10~keV recoil energy) appeared in the nuclear recoil band with some 
coinc\"\i dences between detectors. The corresponding ionization over recoil energy versus 
recoil energy data are plotted in fig.\ref{cesium}.

Nuclear recoil coinc\"\i dences observed between the detectors are probably due to neutrons. 
Such a high background has not been observed in physics runs, this suggests a contamination 
by a californium source of the cesium source holder. This hypothesis 
was confirmed by measurements of the source holder in the low background Ge diode setup placed in the LSM. This blind 
test shows that Edelweiss is indeed sensitive to very low nuclear recoil rates, since detectors are 
able to discriminate about 30 nuclear recoil events among about 10$^5$ events. A WIMP with a mass
of 52~GeV/c$^2$ and a cross section with a nucleon of 7.2$\times$10$^{-6}$~pb, would have yielded
approximately 20 nuclear recoils in the three months of data taking described in~\ref{data}.     

\section{Edelweiss II}

In March 2004, the Edelweiss I experiment has been stopped to allow the installation of 
the second stage Edelweiss II. The aim is an improvement in sensitivity of a factor 100 and reach 
most favored SUSY models. A new low radioactivity cryostat (with a capacity of 50~$\ell$), 
able to receive up to 120 detectors, is being tested in the CRTBT laboratory at Grenoble. 
The first runs will be performed with twenty-one 320~g Ge detectors equipped with NTD heat 
sensor and seven 400~g Ge detectors with NbSi thin film. With an improved polyethylene and 
lead shielding and an outer muon veto, the expected sensitivity is $\sim$~0.002 
evt/kg/day. 

\section*{Note added in proof}

During the completion of this paper, new results from the CDMS experiment have been published~\cite{cdms2}.
They obtained new data with the experiment installed in the Soudan Underground Lab using four 
Ge and two Si detectors. They improve their previous published sensitivity\cite{cdms}, obtained
with the same detectors, by a factor eight .

\end{document}